\documentclass{aip-cp}

\usepackage[numbers]{natbib}
\usepackage{rotating}
\usepackage{graphicx}
\usepackage{verbatim}

\begin{document}

\title{Status and Perspectives of the Search for Eta-Mesic Nuclei}

\author[aff1]{Pawe{\l} Moskal\corref{cor1}}
\eaddress[url]{http://www.aip.org}
\author[aff1]{Magdalena Skurzok\corref{cor2}}
\author[aff2]{Wojciech Krzemie{\'n}\corref{cor3}}

\affil[aff1]{M. Smoluchowski Institute of Physics, Jagiellonian University, 30-348 Cracow, Poland.}
\affil[aff2]{National Centre for Nuclear Research, {\'S}wierk, 05-400 Otwock, Poland.}
\corresp[cor1]{Corresponding author: p.moskal@uj.edu.pl}
\corresp[cor2]{Corresponding author: magdalena.skurzok@if.uj.edu.pl}
\corresp[cor3]{Corresponding author: wojciech.krzemien@ncbj.gov.pl}

\maketitle

\begin{abstract}

In this report the search for $\eta$-mesic nuclei is reviewed. The brief description of the experimental studies is presented with a focus on the possible production of the $\eta$-nucleus bound states for light nuclei like $^{4}\hspace{-0.03cm}\mbox{He}$ and $^{3}\hspace{-0.03cm}\mbox{He}$.


\end{abstract}

\section{INTRODUCTION}

In this article we briefly report on the search for the $\eta$-mesic nuclei. The text includes descriptions from our recent works published in references~\cite{Acta_2016, FewBody_2014, Skurzok_PhD, Krzemien_PhD}. The interested reader is also referred to the recent reviews on mesic nuclei search~\cite{Acta_2016, Machner_2015, Wilkin_2016, Krusche_Wilkin, Kelkar, HaiderLiu, Moskal_FewBody, Moskal_Acta2016}.

\textit{Mesic nucleus} is one of the theoretically predicted and still not experimentally confirmed exotic object. In contrast to already discovered mesonic atom~\cite{Yamazaki_INSRep1996}, where negatively charged pion is trapped in the Coulomb potential of atomic nucleus, this new kind of mesonic matter consists of nucleus bound exclusively via strong interaction with neutral meson such as $\eta$, $\eta'$, $K$, $\omega$. One of the most promising candidates for such state is the $\eta$-mesic nucleus, postulated by Haider and Liu over thirty years ago~\cite{HaiderLiu1}. The coupled-channel studies of the $\pi N \rightarrow \pi N$, $\pi N \rightarrow \pi \pi N$ and $\pi N \rightarrow \eta N$ reactions indicated that in the close-to-threshold region the $\eta$-nucleon interaction is attractive and strong enough to form an $\eta$-nucleus bound system~\cite{BhaleraoLiu}. 

According to the theoretical considerations, the formation of the $\eta$-mesic bound state can only take place if the real part of the $\eta$-nucleus scattering length is negative (attractive nature of the interaction), and the magnitude of the real part is greater than the magnitude of the imaginary part corresponding to the $\eta$ meson absorption:
\begin{equation}
|Re(a_{\eta-nucleus})|>|Im(a_{\eta-nucleus})|.\label{eq:jeden}
\end{equation}

\noindent Initially, it was indicated that due to the relatively small value of the $\eta$N scattering length estimated in eighties, \mbox{$\eta$-mesic} nuclei could be formed only with nuclei having masses greater than 12. However, current researches show that $\eta$-nucleon interaction is considerably stronger than it was expected earlier~\cite{Moskal_habil}. A wide range of possible values of the $\eta N$ scattering length $a_{\eta N}$, 
extracted in some analysis reported in~\cite{Kelkar}, has not excluded the creation of $\eta$-nucleus bound states for a light nuclei such as $^{4}\hspace{-0.03cm}\mbox{He}$, $^{3}\hspace{-0.03cm}\mbox{He}$, T~\cite{Wilkin1,WycechGreen} and even for deuteron~\cite{Green}.

The discovery of $\eta$-mesic bound systems would give unique possibility for better investigation of the elementary meson-nucleon interaction in nuclear medium for low energy region. Moreover, it would provide information about nucleon $N^{*}(1535)$ resonance~\cite{Jido}, about $\eta$ meson properties in nuclear medium~\cite{InoueOset} as well as about the $\eta$ and $\eta'$ meson structure~\cite{BassTom, BassTomek}.
 
The bound states have been searched in many experiments. Nevertheless, none of them gave empirical confirmation of their existence. There are only signals which might be interpreted as an indications of the $\eta$-mesic nuclei. 

\section{EXPERIMENTAL SEARCHES OF ETA-MESIC NUCLEI}

The overview of the measurements carried out in heavy and light nuclei regions one can find in~Ref.~\cite{Skurzok_PhD, Krzemien_PhD}. This section includes the summary of previous experiments and description of currently obtained results.

\subsection{Heavy Nuclei Region}

According to first theoretical predictions $\eta$ meson was more likely to bind to a nuclei with the large number of nucleons, since the attraction of $\eta$ meson could be high enough that allows to form a bound state. Therefore, at the beginning, the experimental searches were concentrated on the heavy nuclei systems. 

First of such experiment devoted to the search for $\eta$-mesic nuclei was performed at BNL~\cite{chrien_bnl} 
with the positively charged pion beam scattered on lithium, carbon, oxygen and aluminium targets ($\pi^{+} + A \rightarrow p + (A-1)_{\eta}$). The missing mass spectra of protons did not show any structure which could indicate the existence of the bound system. However, the BNL experiment was carried out in a region far from the recoilless kinematics that significantly reduced the probability of the possible bound state formation in the ($\pi$, $N$) process. Therefore, a new experiment with pion beam, which is planned at \mbox{J-PARC}~\cite{Fujioka_ActPhysPol2010, Fujioka2} for $(\pi^{-},n)$ reaction will be carried out with the most optimal kinematic conditions.

Another investigations were performed at LAMPF ~\cite{Lampf} in Los Alamos where the $\eta$-mesic $^{18}\hspace{-0.03cm}\mbox{F}$ production was considered based on double charge exchange reaction (DCX). In this picture the bound state is produced via collision of $\pi^{+}$ beam with the neutron inside oxygen target and decays via absorption of the $\eta$ meson on the neutron and, consequently, the emission of $\pi^{-}$. Measured excitation functions also did not reveal peak structure which could be the signature of the $\eta$ mesic nuclei. 

The first evidence for the existence of an $\eta$-mesic bound states was claimed by the LPI group~\cite{Sokol_1999,Sokol_2001} which carried out the measurement of photoproduction processes: (i) $\gamma + ^{12}$\hspace{-0.03cm}$\mbox{C}\rightarrow N + (A-{\eta}) \rightarrow N + \pi^{+}+n+X$ and (ii)~\mbox{$\gamma + ^{12}$\hspace{-0.03cm}$\mbox{C}\rightarrow N + (A-{\eta}) \rightarrow N + p+n+X$}, where $A$ denotes $^{11}\hspace{-0.03cm}\mbox{C}$ or $^{11}\hspace{-0.03cm}\mbox{B}$ nuclei. The invariant mass distribution of the correlated $\pi^{+}n$ pairs for the first process shows a narrow peak structure below the $\eta$ meson production threshold which is in agreement with the theoretical prediction supporting the $\eta$-mesic nuclei formation via $N^{*}(1535)$ resonance excitation and its decay into $\pi$-nucleon pair. The second process was dedicated to search for $\eta$-mesic nuclei through observation of the two-nucleon decay mode arising to the two-nucleon annihilation of $\eta$ meson in the nucleus~\cite{Baskov_2012}. Obtained proton spectra reveal the structure which can also be the indication of the bound system. The upper limit of the photoproduction cross section for both measured reactions was determined and is equal to $10~\mu$b.  

The JINR collaboration performed the search for back-to-back $\pi^{-} p$ pairs related to the $\eta$-mesic bound states in $d$ + $^{12}\hspace{-0.03cm}\mbox{C} \rightarrow \pi^{-}+p + X$ process~\cite{Afanasiev}. An observation of the $\pi^{-} p$ back-to-back correlation as well as the resonance like structure below $\eta$ production threshold could be associated with the two-body $N^{*}$ resonance decay related with formation of an $\eta$-mesic nucleus. However, the experiment needs to be repeated with more intensive beam and the higher spectrometer acceptance.
 
The COSY-GEM measurement of recoil-free $p(^{27}\hspace{-0.03cm}\mbox{Al},^{3}\hspace{-0.03cm}\mbox{He})\pi^{-}p'X$ process results in observation of an enhancement in the missing mass spectrum of the $^{3}\hspace{-0.03cm}\mbox{He}$ which could be interpreted as a signal from $^{25}\hspace{-0.03cm}\mbox{Mg}$-$\eta$ bound state~\cite{Budzanowski,Machner_ActaPhys}. However, it is important to confirm the result with higher statistics. Therefore, the upper limit of the total cross section for the $\eta$-mesic magnesium production was determined and is equal to 0.46$\pm$0.16(stat)$\pm$0.06(syst)~nb.  
 
Another recoil-free experiment dedicated to search for $\eta$-mesic nuclei  was carried out at GSI~\cite{Gillitzer_ActaPhysSlov2006}. The ($d$,$^{3}\hspace{-0.03cm}\mbox{He}$) reaction was measured on $^{7}\hspace{-0.03cm}\mbox{Li}$ and $^{12}\hspace{-0.03cm}\mbox{C}$ targets at GSI Fragment Separator System (FRS). However, till now no final result is published.

\subsection{Light Nuclei Region}

According to the recent investigations~\cite{Kelkar}, the formation of $\eta$-nucleus bound states could also proceed for a light nuclei. In this case absorption is smaller, therefore, the bound states are expected to be narrower in comparison to the heavy nuclei systems what makes them a good candidates for the study of possible binding. 

\begin{figure}[h!]
\centering
\includegraphics[width=6.cm,height=4.2cm]{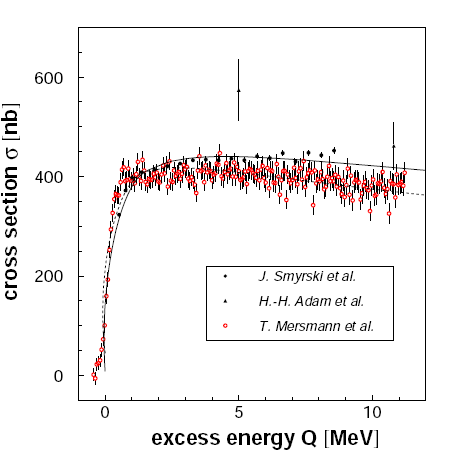} \hspace{0.5cm}
\includegraphics[width=5.8cm,height=4.0cm]{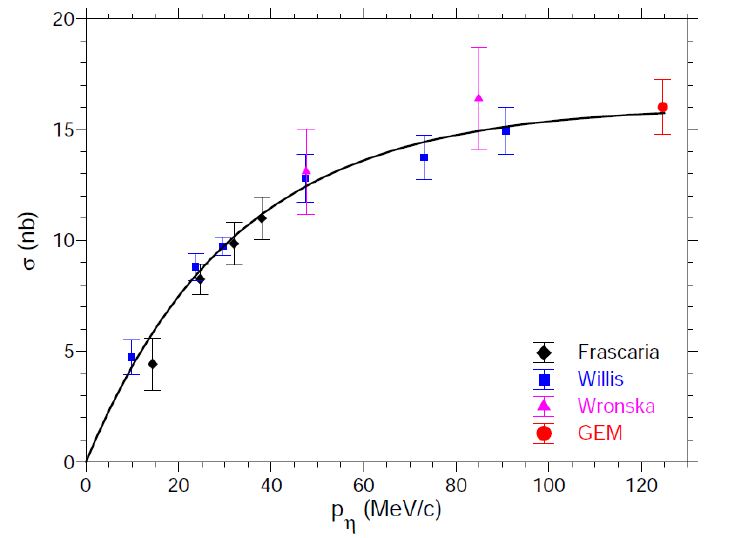}
\vspace{-0.3cm}
\caption{(left) Total cross-section for the $dp\rightarrow$ $^{3}\hspace{-0.03cm}\mbox{He}\eta$ reaction measured with the \mbox{ANKE} (open circles)~\cite{Mersmann} and the \mbox{COSY-11} facilities (closed circles)~\cite{Smyrski1} and (triangles)~\cite{Adam}. Scattering length fit to the ANKE and COSY-11 data is represented with dashed and solid lines, respectively. (right) Total cross-section for the $dd\rightarrow$ $^{4}\hspace{-0.03cm}\mbox{He}\eta$ reaction as a function of CM momentum obtained from the measurements of Frascaria et al.~\cite{Frascaria} (black diamonds), Willis et al.~\cite{Willis} (blue squares), Wro{\'n}ska et al.~\cite{Wronska} (magenta triangles) and Budzanowski et al.~\cite{Budzanowski1} (red circle). The solid line represents a fit in the scattering length approximation. The figure is adopted from~\cite{Machner_ActaPhys}.\label{sigma_3He_4He}}
\end{figure}

\begin{figure}[h!]
\includegraphics[width=8.0cm,height=5.0cm]{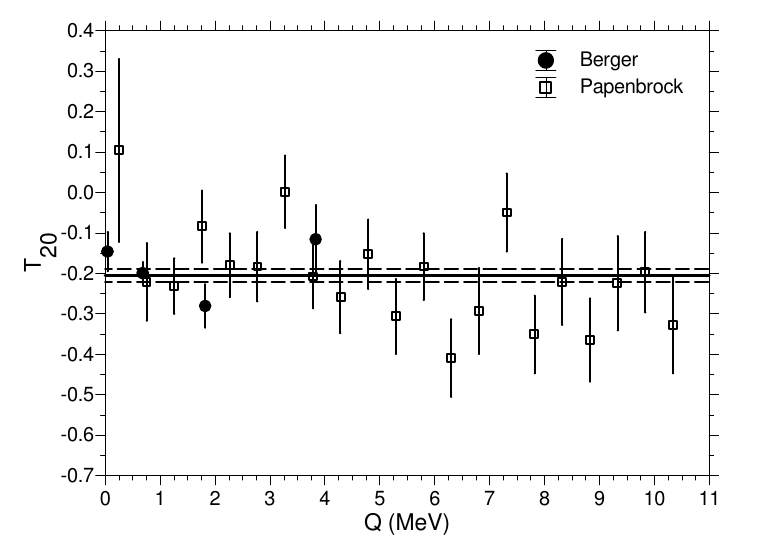} 
\vspace{-0.4cm}
\caption{The angle averaged tensor analysing power $T_{20}$ for the $dp\rightarrow$ $^{3}\hspace{-0.03cm}\mbox{He}\eta$ process as function of the excess energy $Q$. The full dots present the data from~\cite{Berger} while open squares show the data from~\cite{Papenbrock}. Figure is adopted from~\cite{Machner_2015}.\label{ANKE_tensor20}}  
\end{figure}

\begin{figure}[h!]
\centering
\includegraphics[width=8.0cm,height=5.0cm]{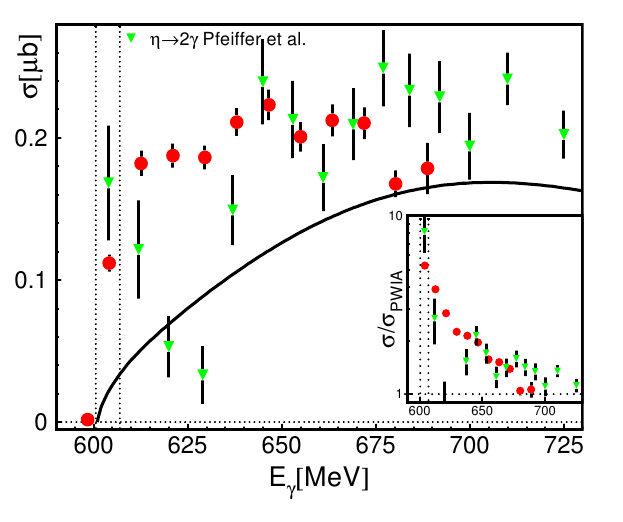}
\vspace{-0.4cm}
\caption{Total cross section for the $\gamma^{3}\hspace{-0.03cm}\mbox{He}\rightarrow$ $^{3}\hspace{-0.03cm}\mbox{He}\eta$ reaction. The green triangles are from~\cite{Pfeiffer} while red dots from~\cite{Pheron}. The two vertical lines indicate the coherent and the break up thresholds. The inserts show the ratio of data and PWIA prediction. Figure is adopted from~\cite{Krusche_Wilkin}.\label{mami_xs}}
\end{figure}

Experimental investigations of the final state interactions (FSI) for $^{3}\hspace{-0.03cm}\mbox{He}\eta$ and $^{4}\hspace{-0.03cm}\mbox{He}\eta$ systems give an observations which may suggest the existence of the bound state. An observed steep rise in the total cross section measured for \mbox{$dp\rightarrow$ $^{3}\hspace{-0.03cm}\mbox{He}\eta$}~\cite{Berger,Mayer,Smyrski1,Mersmann} and $dd\rightarrow$ $^{4}\hspace{-0.03cm}\mbox{He}\eta$~\cite{Machner_ActaPhys,Frascaria,Willis,Wronska} reactions (Fig.~\ref{sigma_3He_4He}), as well as the small and constant value of measured tensor analysing power $T_{20}$ corresponding to very strong variation of the $s$-wave amplitude for $dp\rightarrow$ $^{3}\hspace{-0.03cm}\mbox{He}\eta$ process~\cite{Berger,Smyrski1,Mersmann,Papenbrock} (Fig.~\ref{ANKE_tensor20}) can be the strong evidence for the existence of a pole in the $^{3}\hspace{-0.03cm}\mbox{He}\eta$ scattering matrix which can be associated with the possible \mbox{$\eta$-mesic} nucleus. Another argument for the $^{3}\hspace{-0.03cm}\mbox{He}$-$\eta$ bound state existence is fact that the rise of the cross section above threshold is independent of the initial channel. The total cross section measured for the photoproduction $\gamma^{3}\hspace{-0.03cm}\mbox{He}\rightarrow$ $^{3}\hspace{-0.03cm}\mbox{He}\eta$~\cite{Pfeiffer,Pheron} reaction (Fig.~\ref{mami_xs}) shows the similar behaviour like those obtained for $^{3}\hspace{-0.03cm}\mbox{He}\eta$ hadron production~\cite{Smyrski1,Mersmann}.

\vspace{-0.3cm}

\vspace{-0.2cm}

\newpage

The first direct experimental sign of a light $\eta$-nucleus bound states was reported by the TAPS group~\cite{Pfeiffer} for the $\eta$ photoproduction process $\gamma^{3}$\hspace{-0.03cm}$\mbox{He}\rightarrow \pi^{0}pX$. An~enhancement just below the threshold of the $\gamma^{3}\hspace{-0.03cm}\mbox{He}\rightarrow$ $^{3}\hspace{-0.03cm}\mbox{He}\eta$ reaction was observed in the spectrum being the difference between excitation curves for $\pi^{0}$-proton opening angles of~$170^{\circ}-180^{\circ}$ and $150^{\circ}-170^{\circ}$ in~the~center-of-mass frame. Obtained result was in agreement with the scheme according to which {$^{3}\hspace{-0.03cm}\mbox{He}$-$\eta$} bound state is created and the $\eta$ meson captured by one of~nucleons inside helium forms an intermediate $N^{*}(1535)$ resonance which decays into pion-nucleon pair. However, the later measurement carried out with much higher statistics~\cite{Pheron} shows that the structure observed in the $\pi^{0}$-$p$ excitation function is an artefact of the complicated background behaviour. 

The search for the direct signal from $\eta$-mesic $^{3}\hspace{-0.03cm}\mbox{He}$ was also performed by COSY-11~\mbox{~\cite{Moskal_2010,SmyMosKrze1,Krzemien1,Smyrski2,Smyrski3}} and COSY-TOF~\cite{Gillitzer_ActaPhysSlov2006} collaborations. The excitation functions measured by COSY-11 group for the $dp\rightarrow ppp\pi^{-}$ and $dp\rightarrow$ $^{3}\hspace{-0.03cm}\mbox{He}\pi^{0}$ reactions in the vicinity of the $\eta$ production threshold allowed to establish the upper limits of the total cross section to about 270~nb and 70~nb, respectively. 

In 2008 and 2010 WASA-at-COSY collaboration performed the measurements dedicated to search for the $^{4}\hspace{-0.03cm}\mbox{He}$-$\eta$ bound states in deuteron-deuteron fusion reaction. The $\eta$-mesic nuclei was searched via studying of excitation function for the $dd\rightarrow$ $^{3}\hspace{-0.03cm}\mbox{He} p \pi{}^{-}$~\cite{Acta_2016,Krzemien_PhD,BS_WASA} (2008 and 2010) and $dd\rightarrow$ $^{3}\hspace{-0.03cm}\mbox{He} n \pi{}^{0}$~\cite{Acta_2016,Skurzok_PhD} (2010) reactions near the $^{4}\hspace{-0.03cm}\mbox{He}\eta$ threshold. The measurements were carried out with the beam momentum changing slowly and continuously around the $\eta$ production threshold in each of acceleration cycle. In the first experiment, the beam momentum interval corresponded to the range of excess energy $Q$ from about -51~MeV to 22~MeV, while in the second was extended corresponding to the excess energy $Q\in(-70,30)$~MeV. 

\begin{figure}[h!]
\includegraphics[width=8.0cm,height=5.0cm]{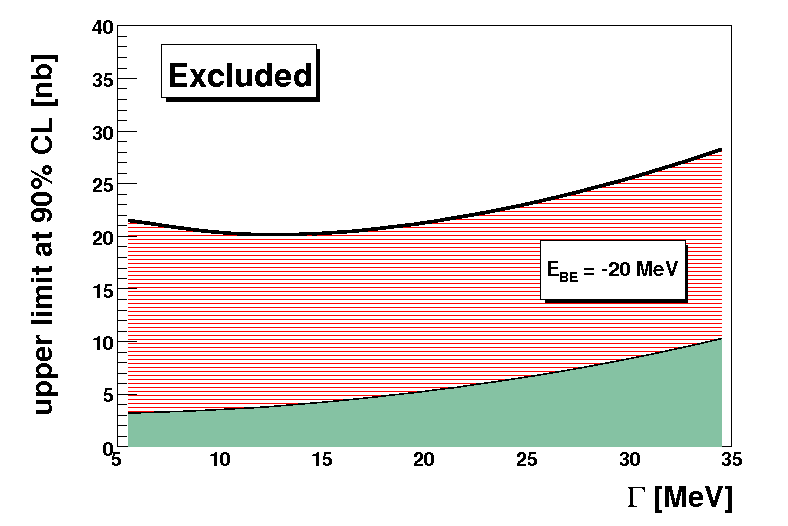} 
\vspace{-0.4cm}
\caption{Upper limit at 90\% confidence level of the cross section for formation of the $^{4}\hspace{-0.03cm}\mbox{He}$-$\eta$ bound state and its decay via the $dd\rightarrow$ $(^{4}\hspace{-0.03cm}\mbox{He}$-$\eta)_{bound}$ $\rightarrow$ $^{3}\hspace{-0.03cm}\mbox{He} p \pi{}^{-}$ reaction as a function of the width of the bound state. The binding energy was set to 20~MeV. The green area at the bottom represents the systematic uncertainties. Figure is adopted from~\cite{BS_WASA}.\label{Wojtek}}  
\end{figure}

Excitation function obtained for the $dd\rightarrow$ $^{3}\hspace{-0.03cm}\mbox{He} p \pi{}^{-}$ reaction measured in earlier experiment does not show the resonance like structure which could be interpreted as a signature of \mbox{$\eta$-mesic} $^{4}\hspace{-0.03cm}\mbox{He}$ bound state~\cite{Krzemien_PhD,BS_WASA}. The upper limit for the cross-section for the bound state formation and decay in the $dd\rightarrow$ $(^{4}\hspace{-0.03cm}\mbox{He}$-$\eta)_{bound}$ $\rightarrow$ $^{3}\hspace{-0.03cm}\mbox{He} p \pi{}^{-}$  process was determined at the 90\% confidence level by the fitting the excitation curve with Breit-Wigner function (signal) with fixed binding energy and width combined with second order polynomial (background). It was noted that the upper limit depends mainly on the width of the bound state and only slightly on the binding energy. The obtained result is presented in Fig.~\ref{Wojtek} for binding energy 20~MeV. It varies from 20~nb to 27~nb as the width of the bound state varies from 5~MeV to 35~MeV. 

\begin{figure}[h!]
\centering
\includegraphics[width=6.5cm,height=4.5cm]{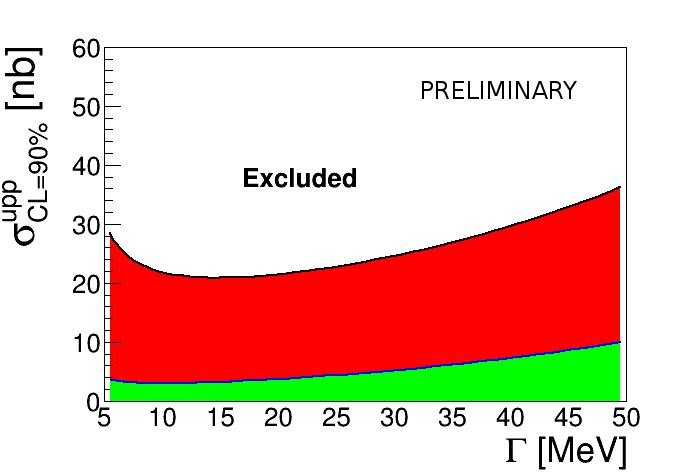}
\includegraphics[width=6.5cm,height=4.5cm]{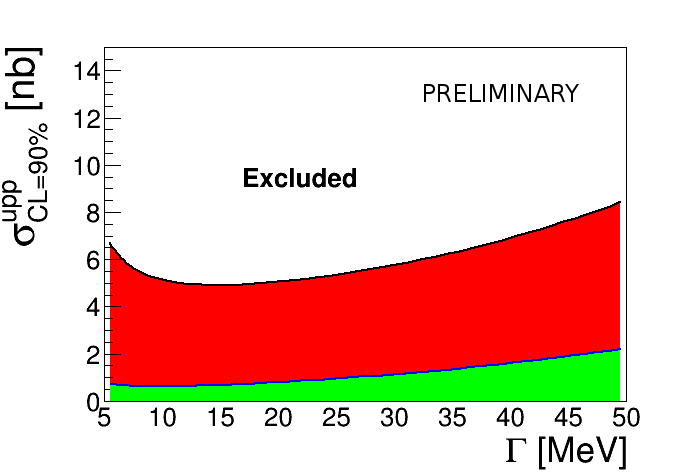}
\caption{Preliminary upper limit of the total cross-section for $dd\rightarrow(^{4}\hspace{-0.03cm}\mbox{He}$-$\eta)_{bound}\rightarrow$ $^{3}\hspace{-0.03cm}\mbox{He} n \pi{}^{0}$ (left panel) and $dd\rightarrow(^{4}\hspace{-0.03cm}\mbox{He}$-$\eta)_{bound}\rightarrow$ $^{3}\hspace{-0.03cm}\mbox{He} p \pi{}^{-}$ (right panel) reaction as a function of the width of the bound state. The binding energy was set to 30~MeV. The green areas denote the systematic uncertainties~\cite{Skurzok_PhD}. The figure is adapted from Ref.~\cite{Acta_2016}.~\label{Result_sigma_upp}}  
\end{figure}

Experiment performed in 2010 allowed to collect the data with about ten times higher statistics. The excitation functions determined for $dd\rightarrow$ $^{3}\hspace{-0.03cm}\mbox{He} p \pi{}^{-}$ and $dd\rightarrow$ $^{3}\hspace{-0.03cm}\mbox{He} n \pi{}^{0}$ processes, however do not reveal any direct narrow structure which could be signature of the bound state with width less than 50~MeV. So far, preliminary upper limit of the total cross section for the $\eta$-mesic $^{4}\hspace{-0.03cm}\mbox{He}$ formation and decay was estimated for both of reactions (Fig.~\ref{Result_sigma_upp}). In case of $dd\rightarrow(^{4}\hspace{-0.03cm}\mbox{He}$-$\eta)_{bound}\rightarrow$ $^{3}\hspace{-0.03cm}\mbox{He} p \pi{}^{-}$ reaction we achieved a sensitivity of the cross section of the order of few nb which is about four times better in comparison with the result obtained from 2008 data~\cite{BS_WASA}. The obtained upper limit value does not exclude the theoretically estimated value $\sigma_{tot}=$4.5~nb~\cite{WycechKrzemien}.
The excitation function for the $dd\rightarrow(^{4}\hspace{-0.03cm}\mbox{He}$-$\eta)_{bound}\rightarrow$ $^{3}\hspace{-0.03cm}\mbox{He} n \pi{}^{0}$ reaction was for the first time determined experimentally. The obtained upper limit is here by factor of five larger than predicted value therefore, one can conclude, that the measurement does not exclude the existence of bound state in this process either~\cite{Kelkar_2015_new}. 

The excitation functions determined for both of reactions are a subject of interpretation of few theoretical groups~\cite{Kelkar_Acta2016,Miyatani_Acta2016} with respect to very wide \mbox{$(^{4}\hspace{-0.03cm}\mbox{He}$-$\eta)_{bound}$} or \mbox{$^{3}\hspace{-0.03cm}\mbox{He}$-$N^{*}$} bound system~\cite{Kelkar_2015_new}.

\section{PERSPECTIVES}

\indent In May 2014, the WASA-at-COSY collaboration performed the measurement dedicated to search for $^{3}\hspace{-0.03cm}\mbox{He}$-$\eta$ bound states~\cite{Proposal_new} in processes corresponding to the three mechanisms: (i) absorption of the $\eta$ meson by one of the nucleons, which subsequently decays into $N^{*}$-$\pi$ pair e.g.: $pd \rightarrow$ ($^{3}\hspace{-0.03cm}\mbox{He}$-$\eta)_{bound} \rightarrow$ $p p p \pi{}^{-}$ , (ii) decay of the $\eta$ -meson while it is still "orbiting" around a nucleus e.g.: $pd \rightarrow$ ($^{3}\hspace{-0.03cm}\mbox{He}$-$\eta)_{bound} \rightarrow$ $^{3}\hspace{-0.03cm}\mbox{He} 2\gamma$ reactions and (iii) $\eta$ meson absorption by few nucleons e.g.: $pd \rightarrow$ ($^{3}\hspace{-0.03cm}\mbox{He}$-$\eta)_{bound} \rightarrow$ $ppn$. Almost two weeks of measurement with an high average luminosity (6$\cdot10^{30}$ cm$^{-2}$ s$^{-1}$) brought a world largest data sample for $^{3}\hspace{-0.03cm}\mbox{He}$-$\eta$. The data analysis is ongoing. 

Also another international collaborations carry out experimental searches for $\eta$ and $\eta'$- mesic bound states. At J-PARC the search for $\eta$-mesic nuclei is going to be performed in pion induced reaction~\cite{Fujioka2}, while the GSI group for the first time have performed the search for $\eta'$ - mesic bound systems in inclusive measurement of the $^{12}\hspace{-0.03cm}\mbox{C}$(p,d) reaction~\cite{Yoshiki, Fujioka3}. The data analysis is in progress. In parallel to the experimental investigations, several theoretical studies are ongoing~\cite{ Wilkin_2016, Kelkar, BassTom, WycechKrzemien, Hirenzaki_2010, Hirenzaki1, Friedman_2013, Wilkin2, Nagahiro_2013, Niskanen_2015}.

\section{ACKNOWLEDGMENTS}

We acknowledge support by the Foundation for Polish Science - MPD program, co-financed by the European Union within the European Regional Development Fund, by the Polish National Science Center through grants No.~DEC-2013/11/N/ST2/04152, 2011/01/B/ST2/00431, 2011/03/B/ST2/01847 and by the FFE grants of the Forschungszentrum J\"ulich.


\nocite{*}
\bibliographystyle{aipnum-cp}%
\bibliography{sample}%

\end{document}